\documentclass[amsfonts,floats,twocolumn]{revtex4}

\usepackage{graphicx}

\begin{document}

\title{Domain Wall Assisted Magnetic Recording}

\author{A.~Yu.~Dobin}
\email{alexander.y.dobin@seagate.com}
\author{H.~J.~Richter}
\email{hans.j.richter@seagate.com}
\affiliation{Seagate Technology, 47010 Kato Road, Fremont, CA 94538}

\date{May 8, 2005}

\begin{abstract}
Using numerical and analytical micromagnetics we calculated the switching fields and energy barriers
of the composite (exchange spring) magnetic recording media, which consist of layers with high and low
magnetocrystalline anisotropy. We demonstrate that the ultimate potential of the
composite media is realized if the interfacial domain wall fits inside the layers.
The switching occurs via domain wall nucleation, compression in the applied field, de-pinning and propagation
through the hard/soft interface. This domain wall assisted switching results in a significant reduction of the switching field
without substantial decrease of the  for thermal activation energy barrier.
We demonstrate that the Domain Wall Assisted Magnetic Recording (DWAMR)
offers up to a three-fold  areal density gain over conventional single layer recording.
\end{abstract}



\keywords{Ferromagnetic relaxation, damping, spin waves, magnon-magnon scattering}

\maketitle

%

Areal density is one of the most important figures of merits for the information
storage technologies. With the increase of the recording density the
dimensions of the magnetic media grains have to be reduced to maintain the signal-to-noise
ratio.
Reducing grains dimensions results in the decrease of the magnetic energy barrier
which leads to the thermally activated loss of the stored information.
To increase the thermal stability the magnetic anisotropy of the grains can be increased,
however the maximum anisotropy is limited by the write head fields.
The described "trilemma"~\cite{RichterDobin_JMMM04} constitutes the main physical limit that magnetic recording
faces in the attempt to achieve the higher recording densities.

As a means to resolve this problem, Victora and Shen proposed the so-called composite media~\cite{VictoraShen_IEEE05},
which are made of two layers, one of which ("hard") possesses high
magneto-crystalline anisotropy, while the other ("soft") layer has
very low anisotropy. It was shown in
Refs.~\cite{VictoraShen_IEEE05,VictoraShen_IEEEtbp} that for the
optimum exchange coupling across the layers interface
the switching field of such a structure may be
substantially reduced compared to a single layer film.
The concept of exchange spring media, utilizing non-coherent magnetization rotation in
hard/soft multilayers to reduce the switching field,
was independently introduced by Suess~\emph{et al.}~\cite{Vienna_JMMM05,Vienna_APL05}.
Guslienko~\emph{et al.}~\cite{Mryasov_PRB04} analyzed the switching field in dual
layer media taking into account non-uniform magnetization rotation, and
the results were shown~\cite{Vienna_APL05,Mryasov_APL05} to differ from the "two-spin" (rigid layers) approximation.

In this Letter we will show that the ultimate potential of the
composite (exchange spring) media is realized when the interfacial
domain wall fits inside soft and hard layers. We will analyze the
potential of the Domain Wall Assisted Magnetic Recording (DWAMR)
using numerical micromagnetics as well as analytical formulas
derived in Refs.~\cite{Kronmuller_PhysicaB02,Loxley_IEEE01} for
exchange spring magnets.

\begin{figure}[!b]
\includegraphics[width=\linewidth]{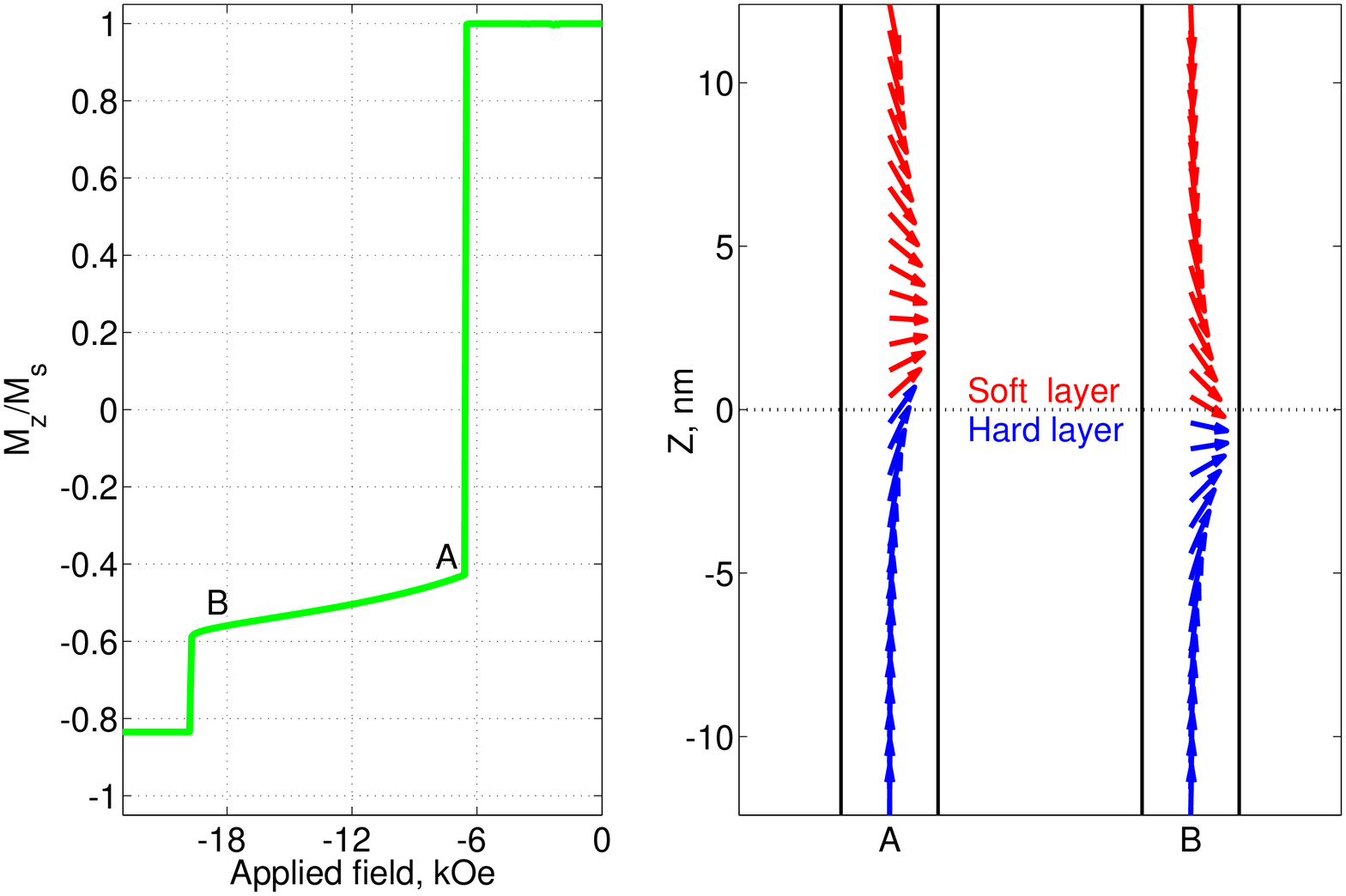}
\caption
  {M-H loop and magnetization configuration for a hard/soft composite film.
  Arrows are drawn for
  every second atomic spin. Distance between spin planes $d_z=2 \AA$.
  Thickness: $t_{soft}=t_{hard}=20$~nm.
  Saturation magnetization: $M_{soft}= 4 M_{hard} = 1600$~emu/cc.
  Anisotropy field: $H_{hard}=200$~kOe, $H_{soft}=6$~kOe.
  Exchange stiffness: $A_{soft}=A_{hard}=1 \cdot 10^{-6}$~erg/cm. Exchange energy density at the interface is the same
  as in bulk, $J_{exc} = 2A/d_z=100$~erg/cm$^2$.}
  \label{Fig:Hysteresis}
\end{figure}

In our micromagnetic simulations the hard and soft layers are
discretized down to atomic scale (0.2 nm) in the vertical (perpendicular to the film plane)
direction. We include the external, anisotropy and vertical
exchange fields, however, for simplicity, we do
not take into account the effects of inter-granular exchange and
demagnetization fields. Therefore our analysis is restricted to a
simplified one-dimensional problem, allowing a better qualitative
understanding of the underlying physics. Although this analysis
cannot be used to quantitatively assess all aspects of recording process,
it is applicable, for
example, for the bit-patterned media, as well as perpendicular
media with demagnetization field fully compensated by the
inter-granular (lateral) exchange field.

In Fig.~\ref{Fig:Hysteresis} an M-H loop and magnetization configurations of a
hard/soft structure are presented. The switching of the upper
(free) part of the soft layer occurs at the applied field close
to the soft layer anisotropy field $H_{soft}=6$~kOe. At this
field the domain wall is pinned near the hard/soft interface
mostly in the soft layer (point A). As the applied field increases the domain wall
in the soft layer is compressed and it penetrates more and more
into the hard layer (point B). Once the applied field reaches $\sim$20~kOe the domain
wall de-pins from the hard/soft interface and propagates through
the entire hard layer. Note, that for the chosen parameters the
hard layer switching field is an order of magnitude less than it's
anisotropy field.

\begin{figure}[!t]
\includegraphics[width=\linewidth]{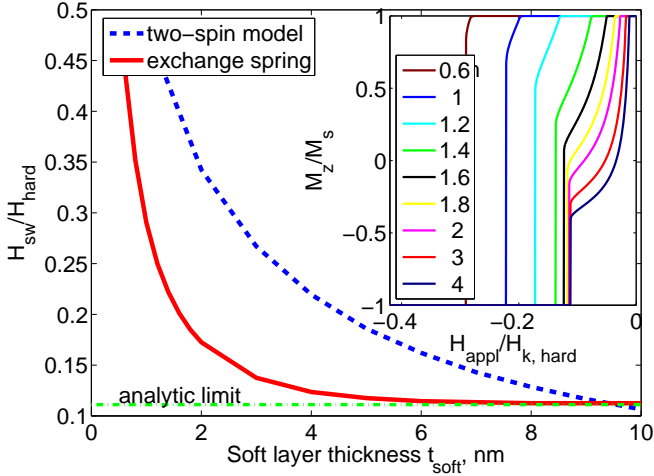}
\caption
  {Switching field vs. soft layer thickness for the exchange spring and two-spin models.
  $H_{soft}=0$, $t_{soft}=t_{hard}=10$~nm, other parameters are the same
  as in Fig.\ref{Fig:Hysteresis}. In the two-spin approximation, the interlayer exchange coupling
  is optimized to provide the lowest switching field. Inset: M-H loops for various soft layer thicknesses.
  }
  \label{Fig:LoopsVsTsoft}
\end{figure}

For infinitely thick hard and soft layers a simple analytic formula for the switching
field was derived and used to study exchange spring magnets~\cite{Kronmuller_PhysicaB02},
as well as domain wall propagation in magnetic wires~\cite{Loxley_IEEE01}:
\begin{eqnarray}\label{Eq:Hsw}
  H_{SW}=\frac{H_{hard}-\gamma H_{soft}}{\left(1+\sqrt{\gamma}\right)^2},
  \quad \gamma \equiv   \frac{M_{soft}A_{soft}}{M_{hard}A_{hard}}.
\end{eqnarray}
Note, that the analytic result $H_{SW}=H_{hard}/4$ was first obtained by Aharoni~\cite{Aharoni_Quarter_PR1960}
in the special case $\gamma=1, \ H_{soft}=0$. This infinite thickness approximation can be applied
to realistic systems with finite layers thicknesses only if the interfacial domain walls fit inside
the hard and (more importantly) soft layers.

\begin{figure}[!t]
\includegraphics[width=\linewidth]{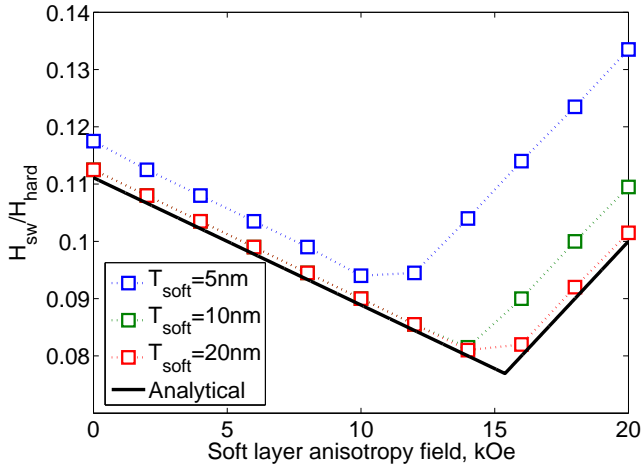}
\caption
  {Switching field dependence on the soft layer anisotropy.
  $M_{soft}=1600$~emu/cc, $M_{hard}=400$~emu/cc;
  $t_{hard}=20$~nm;
  $H_{hard}=200$~kOe;
  $A_{soft}=A_{hard}=1 \cdot 10^{-6}$~erg/cm.
  }
  \label{Fig:HswVsHsoft}
\end{figure}

\begin{figure}[!b]
\includegraphics[width=\linewidth]{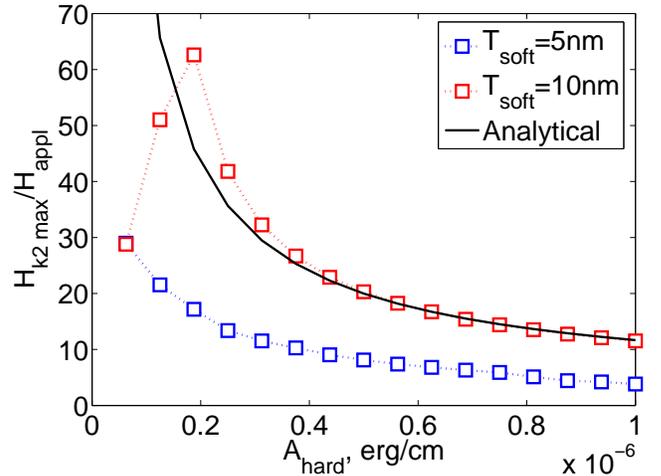}
\caption
  {Maximum hard layer anisotropy field that can be switched by the fixed applied field of
  $H_{appl}=15$~kOe as a function of the hard layer exchange stiffness $A_{hard}$.
  $M_{soft}=1600$~emu/cc, $M_{hard}=400$~emu/cc;
  $t_{hard}=20$~nm;
  $H_{soft}=10$~kOe;
  $A_{soft}=1 \cdot 10^{-6}$~erg/cm.}
  \label{Fig:vsAhard}
\end{figure}

Using infinite continuous layers approximation, similar to that
used in Refs.~\cite{Kronmuller_PhysicaB02,Loxley_IEEE01}, we can estimate the domain wall
thickness in soft or hard layers:
\begin{equation} \label{Eq:DW}
  {\ell_{DW}} \approx \left| \frac{1}{\theta_0}\frac{d \theta}{dz} \right|_{z=0}^{-1} =
    \sqrt{\frac {A \theta_0^2 }{H_{appl}M \left( 1-\cos\theta_0 \right)+K\sin^2\theta_0}},
\end{equation}
where $z$ is the coordinate along the direction perpendicular to the film plane,
$H_{appl}$ is the applied field $K$ and $M$ are anisotropy energy and saturation magnetization of the hard
or soft layer, $\theta_0=\cos^{-1}\left\{ \left( \sqrt{\gamma}-1 \right)/ \left( \sqrt{\gamma}+1 \right) \right\}$
is the magnetization angle at the hard/soft interface $z=0$.
If the anisotropy of the soft material is zero, the domain wall thickness is, of course, infinite
in zero applied field. However, as the applied field increases the domain wall starts to compress
at the hard/soft interface. For the media parameters from Fig.~\ref{Fig:Hysteresis}, the domain wall
thickness at the switching field $H_{appl} \approx 20$~kOe is as small as $\ell_{soft} \approx 2.7$~nm.
For the domain wall assisted switching mechanism to work efficiently, it is required that the interfacial domain wall fits
inside the soft layer, which is demonstrated by simulations in Fig.~\ref{Fig:LoopsVsTsoft}.
Once the soft layer thickness reduces below the domain
wall thickness $\sim 2.7$~nm, the switching field dependence on the soft layer thickness saturates.
The limiting value of switching field is in an excellent agreement with the analytic
prediction of Eq.~(\ref{Eq:Hsw}).
In a striking contrast with these findings,
the switching field with optimized interlayer exchange coupling for the two-spin model depends on the
ratio $M_{soft}t_{soft}/M_{hard}t_{hard}$~\cite{RichterDobin_IEEEtbp}.

The reduction of the switching
field with the soft layer anisotropy, presented in Fig.~\ref{Fig:HswVsHsoft} is another surprising outcome of the domain wall switching mechanism,
which is not predicted by the two-layer approximation.
The switching field reduction with soft layer anisotropy
was first predicted by Hagedorn~\cite{Hagedorn_JAP1970}, who found that in case $\gamma=1$
the minimum switching field $H_{SW}=H_{hard}/5$ is achieved for $H_{soft}=H_{hard}/5$.

Another interesting feature of the domain wall assisted switching is the dependence of the switching field
on the ratio of the exchange stiffness constants in the two layers. In Fig.~\ref{Fig:vsAhard} we present the maximum hard
layer anisotropy field that can be switched by a fixed applied field $H_{appl}=15$~kOe.
For the very small hard layer exchange stiffness the applied field of just 15~kOe can switch the hard layer
with anisotropy as large as 900~kOe!

\begin{figure}[t]
\includegraphics[width=\linewidth]{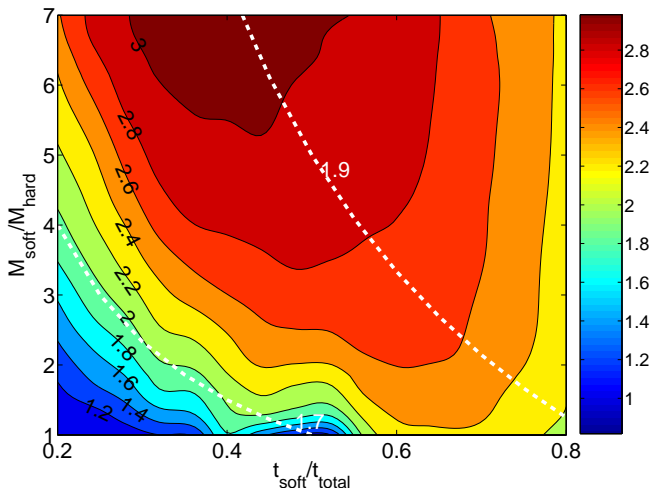}
\caption
  {Optimized DWAMR energy barrier gain (see text for optimization details).
  $H_{appl}=15$~kOe; $t_{tot}=10$~nm; $A_{soft}=2 \cdot 10^{-6}$~erg/cm;
  $A_{hard}=0.5 \cdot 10^{-6}$~erg/cm; $M_{aver} =800$~emu/cc.
  \emph{Dashed lines}: gain calculated within the two-spin approximation, for
  $M_{soft}t_{soft}/M_{hard}t_{hard}$=1 (gain 1.7) and 5 (gain 1.9). }
  \label{Fig:Gain}
\end{figure}


Finally, we assessed the advantage of the DWAMR over the single layer recording using the following
scheme. We calculated the maximum energy barrier for a fixed applied field of $H_{appl}=15$~kOe,
varying the magnetizations, anisotropies and thicknesses of the layers, while keeping constant the total thickness of
the dual layer structure at $t_{tot} \equiv t_{soft}+t_{hard}=10$~nm, and the average magnetization
$M_{aver} \equiv (M_{soft}t_{soft}+M_{hard}t_{hard})/(t_{soft}+t_{hard})=800$~emu/cc.
Energy barriers were calculated using a simplified version of the elastic band approach
suitable for the present 1D problem~\cite{RichterDobin_BarriersTBP}.
In Fig.~\ref{Fig:Gain} the DWAMR energy barrier gain is presented, defined as the ratio of the dual layer energy barrier to the
energy barrier of the equivalent single layer (with thickness $t_{tot}$, saturation magnetization $M_{aver}$ and anisotropy
equal to $H_{appl}$).
In the best conditions DWAMR gain can reach 3, substantially exceeding the maximum value of 2
found in the two-spin model~\cite{VictoraShen_IEEE05,RichterDobin_IEEEtbp}.
The DWAMR gain can be estimated from the following simple arguments, assuming $H_{soft}=0$.
If domain walls fits inside the soft layer,
the switching field (equal to the applied field) does not depend on the layers thicknesses (Eq.~\ref{Eq:Hsw}).
Because the soft layer does not store any energy, the thickness of the soft layer has to be minimized to maximize the gain,
i.e. $t_{soft}=\ell^{DW}_{soft}\approx \sqrt{2A_{soft}/H_{appl}M_{soft}}$ for $\gamma \gg 1$.
On the other hand, the hard layer thickness has to be maximized,
however if it exceeds certain critical thickness $\ell^{DW}_{hard}=4\sqrt{2A_{hard}/H_{hard}M_{hard}}$, the energy barrier
will not increase with thickness anymore~\cite{Braun_PRB1994} saturating at the well-known domain wall energy density (per unit area):
$E^{DW}_{hard}=4\sqrt{A_{hard}H_{hard}M_{hard}/2}$. Maximum gain is then given by
\begin{eqnarray}\label{Eq:MaxGain}
    G_{max}&=&\frac{E^{DW}_{hard}}{H_{appl} \left( M_{hard}\ell^{DW}_{hard}+M_{soft}\ell^{DW}_{soft}\right)/2 }=\\ \nonumber
           &=&\frac{ \left( 1+\sqrt{\gamma} \right)^2}{1+\sqrt{\gamma}(1+\sqrt{\gamma})/4}
\end{eqnarray}
Interestingly, within this simple approximation, maximum gain depends only on the ratio
$\gamma \equiv M_{soft}A_{soft}/M_{hard}A_{hard}$. This dependence has a very flat maximum for $\gamma \gtrsim 20$, reaching
a value of $\approx 4.25$. The accurate numerical optimization (Fig.~\ref{Fig:Gain}) yields a 25\% lower maximum gain value. The
overestimation of the gain in Eq.~(\ref{Eq:MaxGain}) is caused mainly by underestimation of the switching field when Eq.~(\ref{Eq:Hsw})
is used for finite soft layer thickness.

Finally, we want to point out the two most important requirements essential for the high DWAMR gains
and may present significant challenges to the DWAMR implementation.
First, the magnetocrystalline anisotropy of the hard layer has to be as large as $\sim$300~kOe to achieve the highest DWAMR gain,
with relatively low saturation magnetization of $\sim$300~emu/cc.
Presently, no materials with such extremely high anisotropies are known to exist,
thus the future of the composite/exchange spring/DWAMR depends strongly on the discovery of ultra-hard
magnetic materials with good granular structure.
Second, the exchange coupling at the hard/soft layers interface needs to be very large, $\gtrsim$50\% of the bulk exchange energy density
value, which would require an extremely delicate growth process.


\end{document}